\documentclass[aps,prb,twocolumn,notitlepage,superscriptaddress,showpacs]{revtex4-1}

\usepackage{graphicx} 
\usepackage{amsmath}
\usepackage{amsfonts}
\usepackage{mathrsfs}
\usepackage{bm}
\usepackage{color}

\DeclareMathOperator{\im}{Im}

\begin{document}
\title{Finite temperature stability and dimensional 
crossover of exotic superfluidity in lattices}

\author{Miikka O. J. Heikkinen}
\affiliation{COMP Centre of Excellence, Department of Applied Physics,
Aalto University, FI-00076 Aalto, Finland}
\author{Dong-Hee Kim}
\affiliation{COMP Centre of Excellence, Department of Applied Physics,
Aalto University, FI-00076 Aalto, Finland}
\affiliation{Department of Physics and Photon Science, Gwangju Institute of Science and Technology, Gwangju 500-712, Korea}
\author{P\"{a}ivi T\"{o}rm\"{a}}
\email{paivi.torma@aalto.fi}
\affiliation{COMP Centre of Excellence, Department of Applied Physics,
Aalto University, FI-00076 Aalto, Finland}
\affiliation{Kavli Institute for Theoretical Physics, University of California, Santa Barbara, California
93106-4030, USA}

\begin{abstract}
We investigate exotic paired states of spin-imbalanced
Fermi gases
in anisotropic lattices, tuning the dimension
between one and three.
We calculate
the finite temperature phase diagram of the system
using real-space dynamical mean-field theory in combination with 
the quantum Monte Carlo method.
We find that regardless of the intermediate dimensions examined,
the Fulde-Ferrell-Larkin-Ovchinnikov (FFLO) 
state survives to reach about one third of the BCS 
critical temperature of the spin-density balanced case. 
We show how the gapless nature of the state found is reflected in the local spectral function.
While the FFLO state is found at a wide range of polarizations at low 
temperatures across the dimensional crossover,
with increasing temperature we find out strongly dimensionality-dependent 
melting characteristics of shell structures related to 
harmonic confinement.
Moreover, we show that intermediate dimension can help to stabilize an 
extremely uniform finite temperature FFLO state despite 
the presence of harmonic confinement.
\end{abstract}


\maketitle

Fermion pairing in the presence of spin-density imbalance has been a fundamental issue 
in many strongly correlated systems of 
many fields ranging from 
superconductors to ultracold atomic gases and neutron stars \cite{casalbuoni2004}.
While a large magnetic field is detrimental to BCS
superconductivity \cite{Clogston1962,Chandra1962},
it has been predicted that a more exotic pairing mechanism would maintain Cooper pairs coexisting 
with finite spin-density imbalance. The Fulde-Ferrell-Larkin-Ovchinnikov (FFLO) state 
suggests a promising scenario where, with spin polarization, Cooper pairs carry nonzero 
center-of-mass momentum, exhibiting a spatially 
oscillating order parameter \cite{FF,LO}. In three-dimensional 
continuum, the previous mean-field studies predicted only a tiny FFLO area in the phase diagram 
\cite{Sheehy2006}. Indeed, the FFLO state remains elusive in experiments in spite of 
indirect evidence of its existence 
reported in several fields \cite{Radovan2003,Bianchi2003,Kumagai2006,Correa2007,Hulet2010}. 
However, the stability of
the FFLO state has been suggested to depend rather sensitively on system 
settings, such as the presence of lattices 
\cite{timo3,Loh2010,Batrouni2008,Wolak2012},
the dimensionality
\cite{Gora2005,orso2007,hu2007,parish2007,liu2008,zhao2008,Wolak2010}, 
and the trap aspect ratio \cite{Kim2011}.
In particular, it has been anticipated that the FFLO signature
would be much more visible in a dimensional crossover regime between one-dimensional (1D) and
three-dimensional (3D) systems where the strong 1D-FFLO character at zero temperature could be further
stabilized by the long-range order supported by higher dimensions.

In this paper, we provide finite temperature phase diagrams of spin-polarized Fermi gases 
in lattices of intermediate dimensions in a 1D-3D crossover regime. 
Intermediate dimensions are
accessible in ultracold atomic gases by controlling optical lattices as realized for
weakly coupled 1D tubes \cite{Hulet2010} and chains \cite{Greif2012}.
Previous works on the FFLO state in the dimensional crossover regime are done at the mean-field
level for coupled tubes \cite{parish2007,Sun2012} or at zero temperature for two-leg ladders
\cite{Feigun2009} and lattices \cite{Kim2012}. 
However, finite temperature effects are of fundamental importance and  
need to be fully understood to form a precise picture on the observability of exotic 
paired states, especially regarding shell structures of different phases that occur because of
the overall harmonic trap present in ultracold gas experiments. While at
close to 1D, the scaling of critical temperature was studied by the effective field theory 
\cite{zhao2008}, to our knowledge the finite temperature phase diagram in the intermediate 
dimension has not been systematically approached so far. By employing a real-space variant 
of dynamical mean-field theory (DMFT) \cite{DMFTreview,Snoek2011,Kim2011}
with continuous-time auxiliary-field quantum Monte Carlo method \cite{Gull2008,Gull2011,Koga2010},
we explicitly consider the effects of local quantum fluctuations 
at finite temperatures beyond the mean-field level, and the presence of a trap potential 
which is essential to the shell structures.

\begin{figure*}[ht!]
\includegraphics[width=\textwidth]{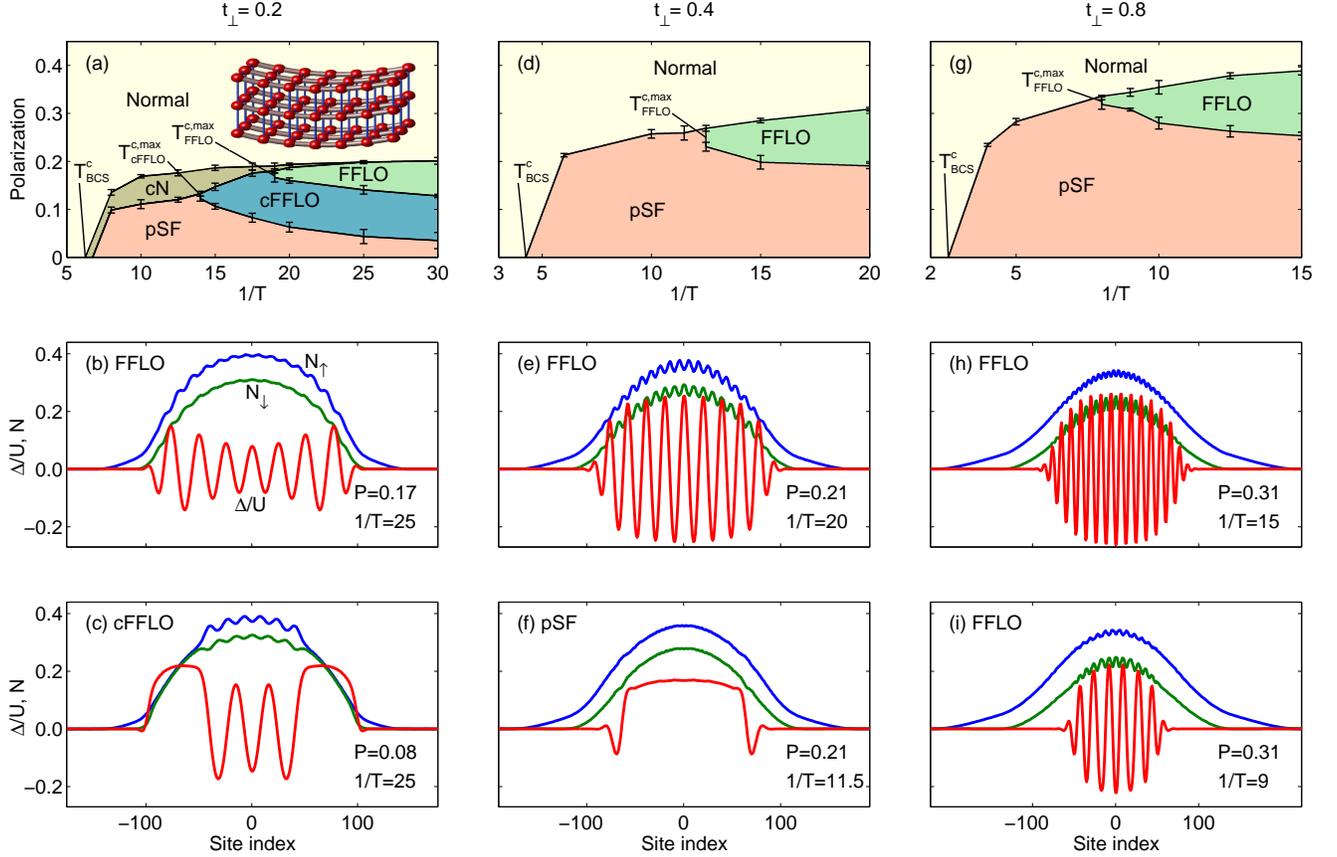}
\caption{\label{fig:phasediags}
The phase diagram of the spin-polarized Fermi gas for different dimensionalities
with two representative density and order parameter profiles 
along the trapped axis for each value of $t_\perp$.
In the phase diagrams 
the notation Normal
stands for normal state and pSF for polarized superfluid 
(including a balanced SF as a special case)
while
cN refers to a shell structure where the system is in the normal state in the middle of the trap
with polarized superfluid or the FFLO state on the edges.
Similarly, cFFLO refers to the FFLO state in the middle of the trap
and polarized superfluid on the edges.
Each errorbar is determined by the closest well-converged simulation
on each side of the phase boundary while the boundary itself
is given by the mean of these two points.
(a) The phase diagram for quasi-1D lattice with $t_\perp=0.2$ 
($U=-2.97$)
with (b) the FFLO state and (c) the cFFLO state.
(d) The phase diagram for an intermediate interchain hopping $t_\perp=0.4$ 
($U=-4.44$)
with
(e) the FFLO state melting to (f) polarized superfluid phase at constant polarization.
(g) The phase diagram for a quasi-3D geometry with $t_\perp=0.8$
($U=-6.83$).
Panels (h) and (i) demonstrate how the FFLO state is affected by the increasing temperature.
The inset of panel (a) is a schematic of the system geometry. 
All energies and temperatures are in units of $t_\parallel = 1$.}
\end{figure*}

We consider a trapped, attractively interacting two-component Fermi gas in an optical lattice
of 1D chains which are coupled to form an anisotropic cubic lattice (see the inset of Fig.~\ref{fig:phasediags}(a)). 
For deep lattice potentials, this system is described by the Hubbard Hamiltonian
\begin{eqnarray*}
\mathcal{H} = && -t_\parallel \sum_{il\sigma} 
(c^\dagger_{il\sigma}c^{}_{(i+1)l\sigma} + \mathrm{h.c.}) 
-t_\perp \sum_{\langle l l^\prime \rangle} \sum_{i\sigma}
c^\dagger_{il\sigma}c^{}_{il^\prime\sigma} \\
&& + U \sum_{il} \hat{n}_{il\uparrow} \hat{n}_{il\downarrow} 
+ \sum_{il\sigma} (V_i - \mu_\sigma) \hat{n}_{il\sigma}.
\end{eqnarray*}
Here $\sigma=\uparrow,\downarrow$ denotes the (pseudo) spin state, 
while $l$ and $i$ stand for the chain and the lattice site
within the chain, respectively. The fermionic annihilation and creation 
operators are $c^{}_{il\sigma}$ and $c^\dagger_{il\sigma}$,  and $\hat{n}_{il\downarrow}$ 
is the density operator. 
We consider a harmonic potential $V_i=\omega^2_\parallel i^2/2$ along the chains. 
The model is parameterized by the interaction strength $U$,
the trap frequency $\omega_\parallel$, 
the spin-dependent chemical potential $\mu_\sigma$ and the intra- and
interchain hoppings $t_\parallel$ and $t_\perp$, respectively.
All energies and temperatures 
are in units of $t_\parallel$, and we set $t_\parallel = 1$.
Varying the interchain hopping $t_\perp$ from zero to one the system
undergoes a dimensional crossover from a collection of 1D chains to a 3D (cubic) lattice.
In the calculations, the trap frequency is set to $\omega_\parallel=1.1\times 10^{-2}$.
The chemical potentials are varied to control the 
polarization 
$P=(N_\uparrow-N_\downarrow)/(N_\uparrow+N_\downarrow)$
while keeping the total particle number 
constant
at $N_\uparrow+N_\downarrow\approx 100$ per each chain.
The interaction strength $U$ is chosen for each value of $t_\perp$ to correspond 
to the lattice equivalent of the unitarity limit \cite{Burovski2006}, and thus,
the interaction strength is fixed by a fundamental two-body 
property.\cite{footnote1}
A further investigation of an optimal interaction strength for the realization
of the FFLO state remains beyond the present work.

In DMFT, the self-energy of the system is taken as site diagonal, 
i.e. $\bm{\Sigma}_{il,i'l'}(i\omega_n)=\delta_{i,i'}\delta_{l,l'}\bm{\Sigma}_{i,l}(i\omega_n)$.
We consider the system and all physical quantities 
to be homogeneous 
in the interchain direction, 
and therefore, the self-energy becomes independent of the chain index $l$.
Thus, the Green's function of the system can be written as
\begin{equation*}
[\bm{G}^{-1}(\bm{k}_\perp;i\omega_n)]_{ij} =
[\bm{G}^0_\parallel(i\omega_n)]^{-1}_{ij}-[\epsilon_{\bm{k}_\perp} \bm{\sigma}_3 
+\bm{\Sigma}_{i}(i\omega_n) ] \delta_{ij} 
\end{equation*}
in which $G^0_\parallel$ is the noninteracting Green's function of a single chain,
$\omega_n$ is the Matsubara frequency and $\bm{\sigma}$ is the Pauli matrix.
The transverse kinetic term
is given by the dispersion $\epsilon_{\bm{k}_\perp} \equiv -2t_\perp (\cos k_x + \cos k_y)$
with the transverse quasimomentum $\bm{k}_\perp=(k_x,k_y)$.
In this notation, the bath Green's function of the DMFT calculations is given as
$[\bm{\mathcal{G}}^0_{i}(i\omega_n)]^{-1} = 
[\sum_{\bm{k}_\perp} \bm{G}_{ii}(\bm{k}_\perp;i\omega_n)]^{-1}+\bm{\Sigma}_{i}(i\omega_n)$.
The pairing order is considered within the Nambu formalism, and the order 
parameter is defined as 
$\Delta_i=-\langle c^\dagger_{i,\uparrow} c^\dagger_{i,\downarrow}\rangle$.

We present the phase diagrams for a quasi-1D system with $t_\perp=0.2$, 
a system of intermediate dimensionality with $t_\perp=0.4$, and a quasi-3D system with $t_\perp=0.8$
in Fig.~\ref{fig:phasediags}. 
The order of the phase transitions in Fig.~\ref{fig:phasediags} remains an open question in our study, and it is possible that the phase boundaries are crossovers because of the finite trap potential.
We find that throughout the dimensional crossover, 
the ratio of 
the maximum FFLO critical temperature 
and 
the balanced BCS critical temperature 
is $T^{c,\mathrm{max}}_\mathrm{FFLO}/T^c_\mathrm{BCS}\approx 1/3$. 
Taking the temperature of $0.7~T^{c,max}_\mathrm{FFLO}$ as a reference point, 
we find that the polarization window for the FFLO phase
grows gradually towards the quasi-3D limit from a value of 
$\delta P=0.06$ at $t_\perp=0.2$ to $\delta P=0.10$ at $t_\perp=0.8$.

In the quasi-1D regime we find a superfluid 
order parameter which has its maximum value away from
the center of the trap; this is clearly
visible in the FFLO state of Fig. \ref{fig:phasediags}(b). 
In this regime the FFLO state melts to the shell structure 
of the general type
displayed in Fig.~\ref{fig:phasediags}(c) 
in which there is a polarized superfluid on the edges of the trap and 
an oscillating order parameter at the center, labeled as cFFLO in Fig \ref{fig:phasediags}(a). 
However, above $P=0.18$ the system 
starting from the FFLO state
reaches 
with increasing temperature
the normal state in the middle of the trap (cN)
and not 
the cFFLO shell structure.
The maximum critical temperature of the cFFLO phase is 
$T^{c,\mathrm{max}}_\mathrm{cFFLO}=0.45~T^c_\mathrm{BCS}$.
Below the polarization of $P=0.13$ the cFFLO shell structure melts 
further to a polarized superfluid phase, and above $P=0.13$ to the cN 
shell structure with the normal state in the middle of the trap.

The phase diagrams for systems of intermediate dimensionality ($t_\perp=0.4$) 
and for quasi-3D ($t_\perp=0.8$) are qualitatively similar. In particular, we 
always find the strongest pairing in the middle of the trap similar to a 3D system. 
Comparing Figs.~\ref{fig:phasediags}(e) and \ref{fig:phasediags}(f)
we see how, here in the case of $t_\perp=0.4$,
the FFLO state melts to the polarized superfluid phase
at constant polarization. It is noteworthy, that in the pSF phase the 
edge of the superfluid still exhibits an FFLO-type oscillating order parameter,
which gradually disappears with increasing temperature.
As indicated in Figs.~\ref{fig:phasediags}(d) and \ref{fig:phasediags}(g), 
there is a transition from FFLO to the normal phase above 
the polarizations $P=0.27$ and $P=0.34$ for $t_\perp=0.4$ and $t_\perp=0.8$, respectively.
Figs.~\ref{fig:phasediags}(h) and \ref{fig:phasediags}(i) demonstrate
how the FFLO state responds to an increase in temperature at constant polarization; 
the paired region recedes towards the middle of the trap. Furthermore, the magnitude of the
order parameter decreases while the wavelength of the FFLO oscillation grows; this
observation holds at all dimensionalities. At the low temperature limit of the phase diagrams, 
we find good agreement with the zero temperature results of \cite{Kim2012}.

Throughout the crossover, the lower boundary of the 
FFLO region in the phase diagram
with respect to polarization increases with temperature.
This behavior is essentially explained by the fact that at higher temperatures, 
the polarized superfluid accommodates a larger spin-density imbalance within its thermal excitations.
{\it In the shell structure of a trapped gas this leads to an additional effect which
counteracts the FFLO instability. Namely, the polarization can be redistributed 
towards the BCS-like regions with increasing temperature.}
On the other hand, higher temperatures are less favorable for any pairing effects to take place
and thus the upper boundary of the FFLO region with respect to polarization
is diminished by temperature.
Consequently, throughout the crossover the highest attainable critical 
temperature for FFLO is rather sensitive to polarization.

\begin{figure}
\includegraphics[width=0.44\textwidth]{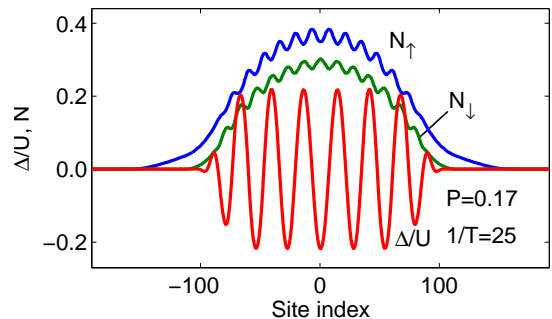}
\caption{\label{fig:twoprofiles}
At the crossing point between quasi-1D and 3D-like behavior, at $t_\perp=0.3$
($U=-3.75$),
we find that the FFLO state (here with $P=0.17$ and $T=0.04$) 
exhibits an exceptionally uniform oscillation amplitude
across the whole system in spite of a density profile varying with the trap potential.}
\end{figure}

Intriguingly, we find that the amplitude of the FFLO order parameter 
is essentially uniform, as shown in Fig.~\ref{fig:twoprofiles},
in a broad polarization and temperature range at the interchain hopping of $t_\perp=0.3$.
This is also the crossing point between 1D-like and 3D-like physics. 
For instance at the temperature of $T=0.05$, such uniform behavior
occurs in a 60~\% interval of the FFLO polarization range in the phase diagram
which is at this temperature $\delta P=0.09$.
This 
suggests that the ``sweet spot'' anticipated in \cite{parish2007}
for observing the textbook FFLO order parameter resides at $t_\perp\approx0.3$.
The uniformity can also be an advantage for observing the state, 
considering probes that rely on strict periodicity of the order parameter.

\begin{figure}
\includegraphics[width=0.48\textwidth]{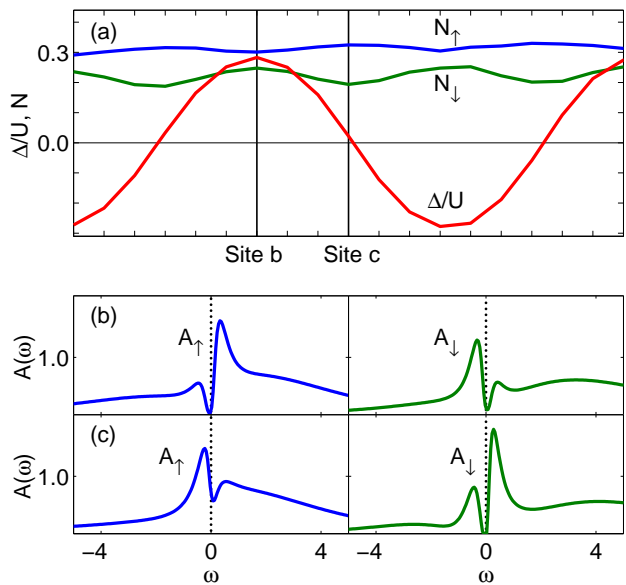}
\caption{\label{fig:specfun}
(a) The FFLO state near the trap center with $t_\perp=0.8$, $T=0.067$ and $P=0.30$.
The local spectral functions of the FFLO state on the 
lattice sites b and c of panel (a) for $\sigma=\uparrow,\downarrow$ 
are plotted in panels (b) and (c), respectively.
The oscillation of the density profile is revealed to affect the most
particles at energies close to the Fermi surface. The gapless nature of
the FFLO state is clearly visible in the majority $(\uparrow)$
component on site c i.e. at the node of the order parameter.
}
\end{figure}

Further characteristics of the FFLO state can be inferred from the
local spectral function plotted in Fig.~\ref{fig:specfun}.
The local spectral function is defined as
$A_{j,\sigma}(\omega)=-2\im G_{jj,\sigma}(i\omega_n\rightarrow \omega+i0^+),$
and can be interpreted as the local density of states.
We use the maximum entropy method to carry out the analytical continuation from the
on-site Green's function obtained from the QMC solver \cite{maxent}.
From the spectral function one can clearly see that the FFLO state is gapless,
and also in this sense the well-known mean-field characterization of the state
remains valid. 
Moreover, we find that only the energy states very close to the Fermi level
contribute to the formation of the FFLO density oscillation.

In conclusion, our work, which incorporates both finite temperature
effects and local quantum fluctuations, 
shows that the FFLO state is significant throughout a dimensional
crossover between 1D and 3D lattices at finite temperatures. The critical temperature
of the FFLO state is approximately one third of the superfluid critical temperature regardless
of the dimensionality, reaching values as high as $T\simeq 0.13~t_\parallel$.
We find that dimensionality has a clear effect on the melting behavior of
the shell structures in a trap, which is essential in distinguishing between the different phases.
Furthermore, we identify $t_\perp=0.3$ as the dimensionality crossover point that provides
the sweet spot of observing a uniform FFLO order parameter despite the harmonic trap confinement.
On a final note, our results confirm that, even at the presence of local quantum fluctuations,
the FFLO state has a wide region of stability in a lattice,
which is in stark contrast to the theoretical predictions in free space where 
the parameter area for the FFLO state is vanishingly small. This gives evidence to the fundamental
role the Fermi surface shape has in stabilizing exotic superfluidity, and 
brings about a significant degree of freedom to future experiments
aimed to realize elusive 
phases of matter 
such as the FFLO state.
From the theoretical point of view it remains an important problem to quantify whether \textit{nonlocal} 
quantum fluctuations play a significant role in the physics of the dimensional crossover and the FFLO state.

\begin{acknowledgments}
This work was supported by the Academy of Finland through 
its Centers of Excellence Programme (2012-2017) 
and under Projects No. 139514, No. 141039, No. 135000,
No. 25174 and No. 263347.
This research was supported in part by the National 
Science Foundation under Grant No. NSF PHY11-25915
and by GIST College's 2013 GUP research fund.
M.O.J.H. acknowledges financial support from the 
Finnish Doctoral Programme in Computational Sciences FICS.
Computing resources were provided by CSC--the Finnish IT 
Centre for Science and the Aalto Science-IT Project. 
\end{acknowledgments}


\begin{thebibliography}{35}
\bibitem[{\citenamefont{{Footnote1}}()}]{footnote1}
\bibinfo{author}{\bibnamefont{{
The hopping ratio $t_\perp = 0.2~t_\parallel$
can be achieved, e.g. with the optical lattice depths of 
$V_\parallel= 5.0~E_r$ and $V_\perp= 10.25~E_r$, 
the trap frequency being $\omega_\parallel= 1.7 \times 10^{-4}~E_r$.
Here, $E_r$ is the recoil energy.
The scattering lengths corresponding to our choice of $U$ are then, e.g.
$a_s=-115~a_0$ and $a_s=-166~a_0$ (with $a_0$ the Bohr radius) for $^{40}$K (with lattice laser wave length $\lambda=738~\mathrm{nm}$)
and $^{6}$Li
(with $\lambda=1064~\mathrm{nm}$), 
respectively.
Similarly the ratio $t_\perp = 0.8~t_\parallel$ can be achieved, e.g. by 
$V_\parallel= 8.0~E_r$ and $V_\perp=8.75~E_r$, 
giving $\omega_\parallel= 6.5 \times 10^{-5}~E_r$.
The scattering lengths would then be $a_s=-92~a_0$
and $a_s=-133~a_0$ for $^{40}$K and $^{6}$Li, respectively.
The values here have been calculated using the harmonic approximation
for each lattice site.
}}}
\end{thebibliography}
\end{document}